\documentclass[12pt]{article} 

\usepackage{amsfonts}
\usepackage{graphicx}
\usepackage{dcolumn}
\usepackage{bm}
\usepackage{amsmath}

\usepackage{graphicx}
\usepackage{dcolumn}
\usepackage{bm}

\newcommand{\dif}{{\rm d}}
\newcommand{\eps}{ { \varepsilon } }
\renewcommand{\phi}{{\varphi}}
\newcommand{\equal}{\buildrel {\rm def} \over {=} }

\newcommand{\vett}[1]{\mathbf{#1}} 

\title 
{ Relaxation times and ergodicity properties \\
  in  a realistic ionic--crystal model,  \\ 
  and the modern form of the FPU problem
}

\author{Andrea Carati\thanks{Dep. Mathematics,
    Universit\`a degli Studi  di  Milano, Via Saldini 50, 20133 Milano
    -- Italy. Corresponding author A. Carati. E--mail address:
    \texttt{andrea.carati@unimi.it}} 
  \and
  Luigi Galgani\footnotemark[1]\addtocounter{footnote}{1}
  \and
  Fabrizio Gangemi\thanks{DMMT, Universit\`a di Brescia,
    Viale Europa 11, 25123 Brescia -- Italy.}
  \and
  Roberto Gangemi\footnotemark[3]
}

\date{\today}

\begin{document}

\maketitle

\begin{abstract}
It is well known that Gibbs' statistical mechanics is not justified
for systems presenting long--range interactions, such as plasmas or
galaxies. In a previous work we considered  a realistic FPU--like
model of an ionic crystal (and thus with long--range interactions), and
showed that it  reproduces the experimental infrared spectra from 1000
K down to 7 K, provided one abandons the Gibbs identification of
temperature in terms of specific kinetic energy, at low temperatures.
Here we investigate such a model  in connection with its
ergodicity properties. The conclusion we reach is that at low
temperatures ergodicity does not occur, and thus    the Gibbs
prescriptions are not dynamically justified, up to geological time scales.
 We finally
give a preliminary result indicating how the  so--called 
``nonclassical'' q-statistics 
show up  in the realistic ionic--crystal model.  How to
formulate a consistent statistical mechanics, with the corresponding
suitable identification of temperature in such nonergodicity conditions,
 remains an open problem, which apparently constitutes  the modern form of 
the FPU problem.
\end{abstract}


\section{Introduction}
The present paper originates from questions of ergodicity type that
naturally arise in connection with a previous paper of ours
\cite{physica} (see also the review \cite{moser}, where the
connections with the FPU (Fermi--Pasta--Ulam) problem were pointed out).
Paper \cite{physica} was devoted to a theoretical computation  of the
infrared spectra of the ionic crystal  LiF (Lithium Fluoride)
in a classical frame. The results, obtained 
through Molecular Dynamics computations of the ionic polarization $\vett
P(t)$ for a realistic model, turned out to reproduce pretty well  the
experimental infrared spectra
 in a range of temperatures from 1000 K down to 7 K. Now, the
relevant quantity that determines the spectra, i.e., the
electric susceptibility $\chi(\omega, T)$ at temperature $T$, is the  Fourier
transform of essentially the time--autocorrelation function of
polarization. On the
other hand correlations are in principle defined as expectations with
respect to a given invariant measure, 
and  so one has to decide which measure should be used in the computations,
 at any temperature $T$.
In practice, however, the expectations were computed 
as time--averages up to
a certain final time, for a little number of random initial data,
extracted in a suitable way
at temperature $T$.

This fact, of substituting time--averages for phase--averages, already
raises some questions, that were indeed posed to us by a referee of
our previous paper.
Namely, does one actually observe a relaxation? And how does the
relaxation time depend on the dynamical quantity considered?  So we
started investigating relaxation times, and actually: 1) the time
required for the vanishing of relevant   time--correlations (usually called
relaxation time in the literature on dynamical systems theory):  
2)  the time required to attain  equipartition of normal--mode energies, 
starting from
exceptional initial data (usually called the relaxation time
in the FPU literature); and finally, 3)  the time required for the
stabilization of the computed spectra, which is  the original question 
raised to us by the referee. In such a way, 
the relaxation times for the three mentioned
cases were found to be, for example at room temperature, of order 1
picosecond, 10 picoseconds and 1 nanosecond respectively.

Such  results naturally raise further questions of ergodic type for the
LiF system. Indeed it is well known  (see \cite{ruelle}
\cite{eckmann}) that for
strongly mixing systems the relaxation times for  smooth
functions do not depend on the function, being essentially related to
 the maximal eigenvalue 
 of the Perron--Frobenius operator of the
dynamical system. So the mentioned results appear to indicate that the
LiF model is not strongly mixing.

On the other hand, there is a  deeper reason for discussing the
ergodicity properties of the model, because  lack of ergodicity at low
temperatures is necessary for its physical consistency.
Indeed, as will be recalled later, in  paper  \cite{physica} 
the agreement   between theoretical
and experimental spectra could be extended to low temperatures only
if one assumes that at such temperatures the relation between specific  
kinetic energy $K/3N$ and temperature $T$ is  different from that  
dictated by the Gibbs statistics, i.e, two  thirds the mean kinetic
energy per particle:
\begin{equation}\label{eq:temperatura}
k_B T = \frac 23 \ \frac KN \ .
\end{equation}
So,  the existence of some dynamical obstructions to
ergodicity is necessary for the physical consistency of the LiF model.

In  the present paper  it will be shown    
that in such a  model ergodicity does not occur up to geological time
scales, so that the  Gibbs prescriptions are not justified 
up to such times. The arguments used clearly indicate that the same
conclusion holds  for all realistic models  of solids, and perhaps 
for the whole domain of atomic physics.

The numerical results on the relaxation times are illustrated in
Section~2,  the relation between temperature and specific energy
dictated by the agreement of experimental and computed spectra
 is shortly recalled in Section~3, the ergodicity properties 
are discussed in Section~4, and some open problems, 
including the possible relevance of ``nonclassical'' q--statistics, 
are shortly discussed  in  Section~5. The conclusions follow in Section~6. 
For details on the model and
on the results already available the reader is referred to paper
\cite{physica}. We are confident, however, that the rather quick
recollection  given below may suffice to make the present
paper self contained.

\section{Numerical results for the relaxation times}
The LiF ionic--crystal   model 
is defined in the standard way of solid state physics. Namely,  we consider a
system of point particles describing the two species of ions (with
opposite charges), in a
``working cell'' containing an even number $N$ of  particles (typically,  
 $N=512$ or 4096), with  periodic boundary
conditions. The interactions among the ions are the Coulomb ones (cared for their
long range feature through the well--known Ewald procedure), plus a
two--body short range phenomenological potential. The latter
 is introduced,  following Born (see \cite{bornhuang}, Chapter I), 
in order to implicitly take into 
account the role
of the electrons, whose degrees of freedom don't show up in the
model. In the same  Born's spirit one also introduces
``effective charges'' which are substituted for the true charges of
the ions in the expression of the mutual Coulomb forces. So the model
is defined by the Hamiltonian
\begin{equation}\label{eq:1}
  H = \sum_{j,s} \frac {\vett p_{j,s}^2}{2m_s} + \sum_{j,j',s,s'}
  V_{s,s'}\big(|\vett x_{j,s} - \vett x_{j',s'}|\big) \ ,
\end{equation}
where $s=1,2$ denotes the ionic species (of  mass $m_s$), while $\vett p_{j,s}$, and 
$\vett x_{j,s}$, are the momentum and position of the
$j$-th ion of the
$s$ species. For the phenomenological potential we choose the 
 Buckingham one, so that the total potential has the form
\begin{equation*}
  V_{s,s'}(r) = a_{s,s'}e^{-b_{s,s'}r} + \frac {c_{s,s'}}{r^6} +
  \frac {e_s^{\mathrm{eff}}e_{s'}^{\mathrm{eff}}}{r}  \ .
\end{equation*}
The values of the constants $ a_{s,s'}$, $b_{s,s'}$, $c_{s,s'}$, and of
the effective charges $e_s^{{\mathrm{eff}}}$ are taken from paper 
\cite{physica}.
Notice that, due to the periodicity conditions, for what concerns the
potential  the sum extends to  all points of the
infinite lattice, although a cutoff of  5 \AA\  was  introduced for
the phenomenological potential.

Having fixed  a lattice step (which we take,  at any temperature,  from
the experimental data), 
and thus a  volume $\mathrm{V}$ of the
working cell,  one can check that, with the chosen values of the
parameters of the potential\footnote{The parameters were chosen by
  optimizing the agreement between experimental and computed spectra
  at room temperature, and were not changed at the other
  temperatures. We leave for a future work the task of a better choice
  of   the parameters, by optimizing the agreement for the whole  set of 
  available  spectra at different  temperatures.}, the system   
 has an equilibrium point, in which the particle
positions form a face--centered cubic lattice. Thus,  at sufficiently low
energies the particles oscillate about their equilibrium positions,  and
the model can be considered as a three--dimensional realistic
variant  of the standard one--dimensional FPU model. 

The numerical investigations were performed by integrating the
equations of motion through the Verlet algorithm with a time step of 2
femtoseconds, for times  up to 2 nanoseconds.

\subsection{Decay--time of correlations, as a function of specific energy}
\begin{figure}  
\begin{center}
    \includegraphics[width = 1.\textwidth]{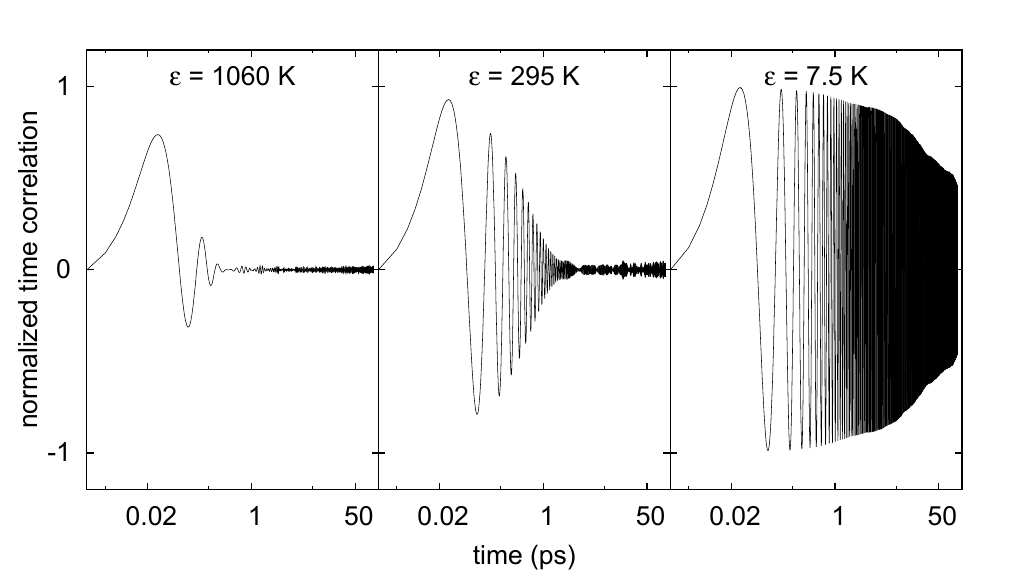}
  \end{center}\
  \caption{Normalized correlation between  polarization and its time derivative
    versus time, for three  different values of the specific energy 
$\varepsilon$ 
    (in Kelvin units). Figure taken from \cite{physica}. N=4096.}
 \label{fig:1}
\end{figure}
With the  aim of studying the relaxation times for relevant observables,
we start considering the ionic polarization $\vett P$, which is defined  by 
\begin{equation*}
  \vett P = \frac 1V \sum_{j,s} e_s\vett  x_{j,s} \ 
\end{equation*}   
(where $e_s$ is the \emph{actual}   charge of the $s$ species of ions), 
and investigate the decay of the correlation
between  ionic polarization and its time--derivative\footnote{In principle one
should consider the correlation tensor involving  products of the
Cartesian components $P_i$ and 
${\dot P}_j$. However, for 
LiF one deals with an isotropic case, and it turns out to be
sufficient  to consider
 only the scalar product $\vett P\cdot \dot{\vett P}$.},
\begin{equation*}
  \overline{\vett  P (t) \cdot  \dot{ \vett P}(0)} \equal
  \lim_{t_{fin}\to+\infty} \frac 1{t_{fin}} \int_0^{t_{fin}}  \vett P (s+t) \cdot    
\dot{\vett P}(s)\, \dif s \ ,
\end{equation*}   
 which is the key quantity
determining  the  spectra of the crystal in the infrared
region (see below). If the system were mixing, one would have $\overline {{\vett
    P}(t) \cdot {\vett P}(0)}\, \to 0$    
as  $t\to\infty$, for almost every initial datum.

The numerical  results found in \cite{physica} are reported in
figure~\ref{fig:1}, in which the time--correlation is given versus time  
for three different
values of the  specific energy $\varepsilon$.
Here, and in the rest of the 
paper,   energy  is defined  as 
the difference between the  value $E$ of the Hamiltonian  $H$
(determined for example by the initial data) 
and  its minimum value
$E_{min}$ (i.e., the minimum $V_{min}$ of the potential 
energy),  so that the specific
energy (per degree of freedom) $\eps$ 
is defined as\footnote{As $E_{min}$ is the  minimum of the potential,   
$\varepsilon$ differs very little from the specific energy
of the normal modes of the quadratic approximation $H_{harm}$ of the Hamiltonian
 about the   equilibrium point. In this way, it becomes
possible to compare our results with those obtained in the
standard FPU model,  which  are usually  expressed as a function of
energy per mode (to which the specific energy  $\varepsilon$ reduces 
in the case  $E_{min}=0$).}
$$
\eps = \frac {E-E_{min}}{3N} \ .
$$
In this paper we chose to express the specific energy $\eps$ in
Kelvin: so, when we say for example that we have $\eps=295$ K we
just mean that $\varepsilon/ k_B = 295$.\footnote{In our
  previous paper, instead of making reference to
 $\varepsilon$ we were making reference to $2K/3N$ ($K$ being the kinetic
  energy), which coincides with $k_BT$ according to the Gibbs
  prescription. Thus its value,   expressed in Kelvin,  was called ``effective
 temperature'' (while it  is called   ``kinetic temperature''
 in the papers of the Tsallis group quoted below). Actually,
  the difference between the effective (or kinetic) temperature and
  the specific energy $\varepsilon$ (expressed in Kelvin) is very
  small
even  at rather large  temperatures  as 1000 K.} 

\begin{figure}  
\begin{center}
  \includegraphics[width = 1.\textwidth]{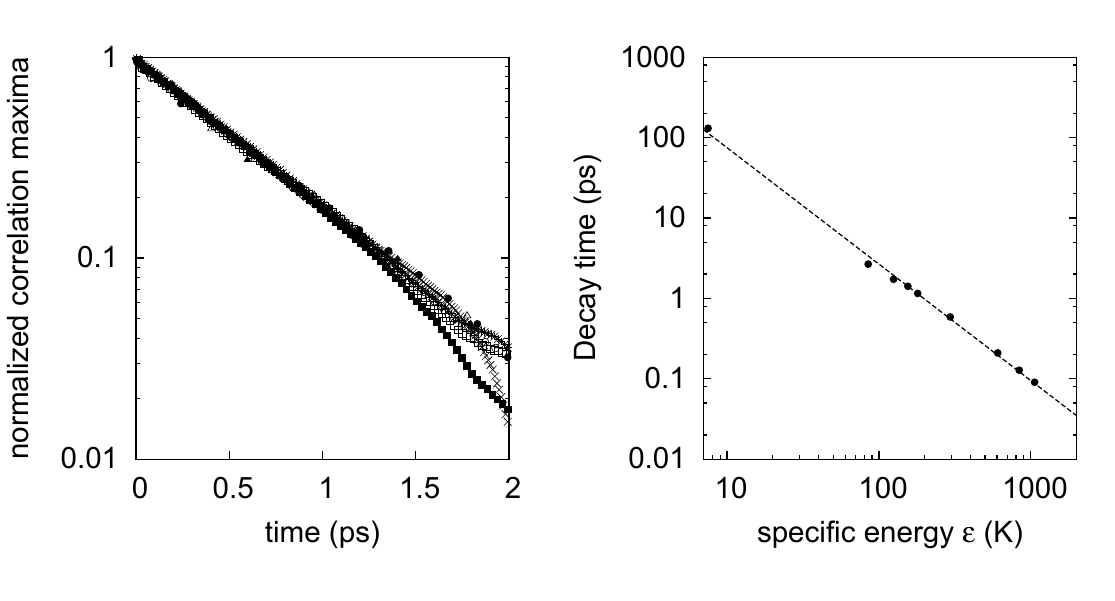}
  \end{center}\
  \caption{Left: the sequences of maxima of the normalized
    correlation function $C^{\eps}$ versus the rescaled time,
    for 7 different values of the specific energy $\eps$.
    Right: Decorrelation time of polarization 
   versus  specific energy $\varepsilon$. $N=4096$.}
 \label{fig:2}
\end{figure}

Every  curve in figure~\ref{fig:1} was obtained as an average over 
10 different initial data, computed up to a finite  time
$t_{fin}=$ 200 picoseconds. As
one sees, the three  correlations seem to vanish for suitable large
times, but the  decay--rate  decreases with the specific energy
$\eps$. To give a quantitative estimate of the
decay--time we proceed as
follows: for each value of $\eps$
we consider the absolute value of the normalized correlation, i.e.,
of the function
$$
C^{\eps}(t) \equal \frac  
{\big| \overline{\vett P (t) \cdot \dot {\vett P}(0)} \big|}
{~~\Big(\,\overline{\rule{0pt}{2.25ex}\vett P^2 (0)}\  
    \overline{ \rule{0pt}{2.25ex}\dot {\vett P}^2(0)}\, \Big)^{1/2} } \ ,
$$
and determine the sequence $C^{\eps}(t_n)$ of its maxima. 
Taking the values at $\eps=295$ K as reference, one finds that, for each
specific energy, there exists a rescaling factor $r(\eps)$ of time, such that the
plots of the sequences of maxima as a function of the rescaled time $t/r(\eps)$
overlap.
 This is exhibited in figure~\ref{fig:2} (left), where the
rescaled sequences for 7 different values of $\eps$ are reported
in  semilog scale. One can notice not only that the different
sequences superpose, but also that they decrease
exponentially\footnote{ At least for not too small values of the
correlation. On the other hand, it is very difficult to reliably
compute  correlations below a certain threshold, because it
would require an exceedingly long stretch of trajectory of the system. So
we discard the tail in our considerations.}
with a decay time $\tau_d(\eps)$ that is estimated to be 0.6 picoseconds at $\eps=295$ K.

The dependence of the decay--time $\tau_d(\eps)$ on   specific
energy is exhibited in figure~\ref{fig:2} (right), where the values found are
reported  in logarithmic scale. As one  sees, 
a straight line with a
slope near to -1.5 gives a good interpolation.
One can thus 
infer  for the decay--time  $\tau_{d}$  a law of the form
\begin{equation}\label{eq:2}
  \tau_d(\eps)=\frac C{\eps^{3/2}} 
\end{equation}
with a suitable $C$.

\subsection{Equipartition time, as a function of specific energy}

\begin{figure}  
  \begin{center}
    \includegraphics[width = 1.\textwidth]{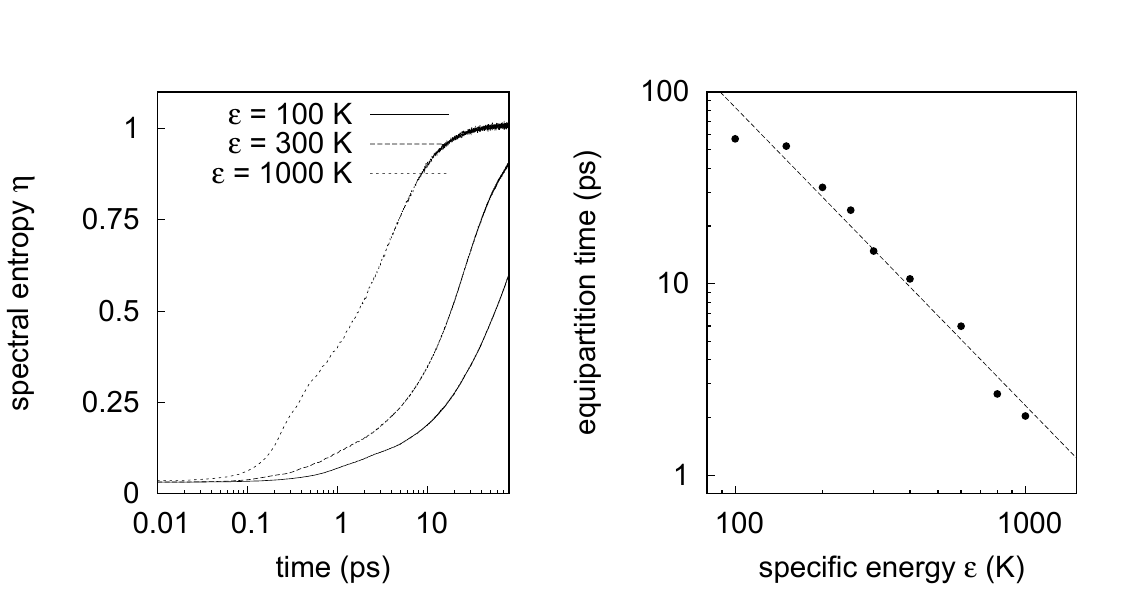}
  \end{center}
  \caption{The approach to equipartition by 0 entropy
    $\eta(t)$. Left: Spectral entropy versus time for
    three different  specific energies. Right: equipartition time $\tau_{eq}$ vs
    specific energy  $\eps$  (points) in logarithmic scale, together with the
    regression line. $N=512$.} 
 \label{fig:4}
\end{figure}

We come now to the equipartition times, i.e., the times needed for
attaining  equipartition  among  the normal--mode energies, 
starting from an
initial condition in which just a  few modes of nearby frequencies 
are excited. Precisely,  we  excite only  the whole packet of
modes (in the number of 15 out of 1536) which have exactly the lowest frequency,  the  modes having 
the same energy but  random phases,

We use two methods. The first one is  standard   in studies 
of the FPU model (see
for example \cite{benettinponno}). 
Denoting  by $E_k$
the energy of the $k$--th normal mode (the definition of normal modes,
and a discussion of their properties will be given   in   Section 4),
one defines the so--called  spectral entropy
$\eta(t)$ by
\begin{equation*}
 \eta(t) = \frac 1{3N} \exp \bigg( -\sum \frac {\overline {E _k}(t)}{E_{harm}} 
\log \frac {\overline{E_k}(t)}{E_{harm}} \bigg ) \ ,
\end{equation*}
where $\overline{ E_k}(t)$ is the time--average of  $E_k$ up to
time $t$, while $E_{harm}$  
is the total harmonic energy (i.e., the sum
of the mode energies).\footnote{ Actualy, $E_{harm}$ too is a function of time. 
However,  as will be shown in Section~4, for the values of $\eps$ we are 
considering it simply presents small oscillations about the value $3N\eps$. 
Here we just neglect such oscillations.} The spectral entropy turns out to be 
very small (of 
order $1/N$)  when just a few modes share the energy $E_{harm}$  ,
 becoming 1 when   equipartition occurs (with $\overline {E_k}=\eps$
 for  $k=1,\ldots,3N$). 
So the attainment  of equipartition corresponds  to finding 
$\eta(t)\simeq  1$ for a large enough $t$. It is usually supposed 
that the attainment  of  equipartition flags the attainment  of  thermal 
equilibrium. This, however, is a delicate point on which we will come back
later. To give a
quantitative estimate of the equipartition times, the following
convention is usually adopted in the FPU literature: 
for any given value of the  specific energy one defines the
equipartition time $\tau_{eq}$ as 
the time  at
which the curve $\eta(t)$  attains a certain value, for example the
value 1/2.

The  results  for the 
spectral entropy are reported 
in figure \ref {fig:4}. In the left panel  we report  the curves   $\eta(t)$  
corresponding to three different 
values of $\varepsilon$. Each such curve was  determined 
by averaging the curves corresponding to five different 
initial data, defined through   a different choice
of the random phases of the excited modes.
\begin{figure}  
\begin{center}
    \includegraphics[width = 1.\textwidth]{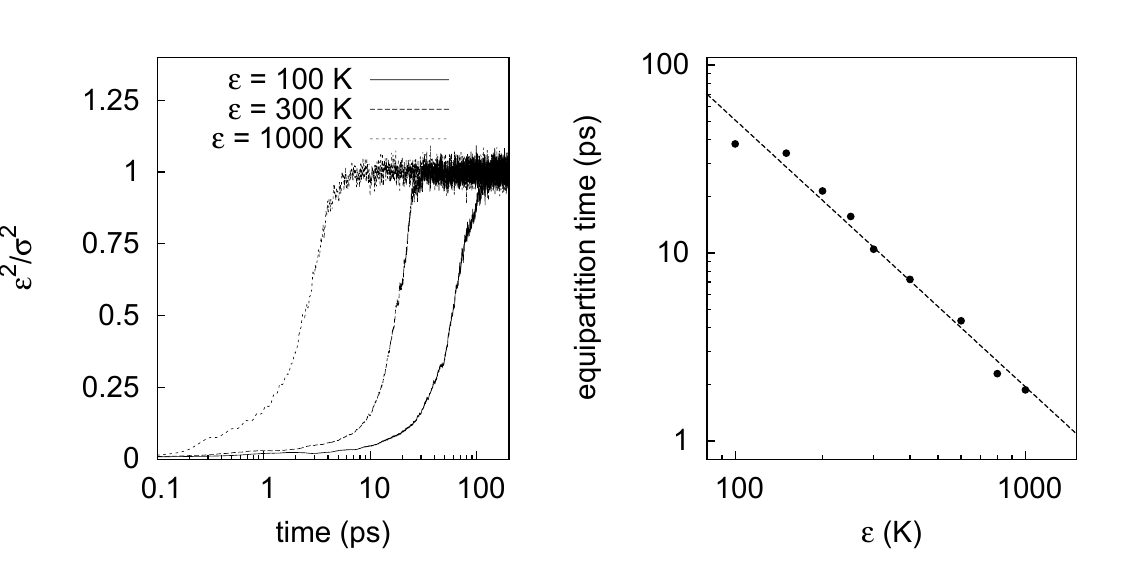} 
  \end{center}\
  \caption{ The approach to equipartition by energy variance
    $\sigma^2$. Left: Inverse of the normalized energy variance versus time. Right:
    equipartition time versus specific energy (points) in logarithmic scale,
     together with the regression line.  $N=512$.} 
 \label{fig:5}
\end{figure}
In agreement with the FPU case, one finds that the
function $\eta$ increases monotonically towards the value 1,  with a
decreasing rate as  $\eps$ is decreased.  The corresponding  values 
of $\tau_{eq}(\varepsilon)$ are reported in
figure~\ref{fig:4}, right  panel, in logarithmic scale. One sees 
that a straight line with a slope not far from -1.5 
 gives a good interpolation of the data.

In order to overcome the conventional 
character of the procedure in which a value such  as  $\eta=1/2$ is
chosen for the spectral entropy  to
flag the attainment  of equipartition,
we used a second method, which was devised by A Giorgilli
\cite{anto}. This is  based on the idea that, by the  attainment of
equipartition of the time--averaged mode energies, the system   might
have actually  attained   also the Maxwell--Bolztmann distribution for
the instantaneous  values of the  mode energies.
If this occurs,  the variance 
$\sigma^2(t)= \frac{1}{3N}\sum_k{\big(E_k(t)\big)^2}-
\left(\frac{1}{3N}\sum_k{E_k(t)}\right)^2$
should converge to $\varepsilon^2$,  and quite naturally  the
relaxation time $\tau_{eq}$  could be defined as
the first  time at which  $\sigma^2(t)/\varepsilon^2$ attains its
asymptotic value, about which it is actually found to subsequently 
oscillate. However, in order to make a consistent comparison with the results
relative to the first method (in which case the value 1 is never
reached) we still define the equipartition
time $\tau_{eq}(\eps)$ as the first time at which $\sigma(t)=2\eps$.

The results for  the second method are better understood if one
reports, as a function of $t$,  the inverse $\eps^2/\sigma^2(t)$  
of the normalized variance, and this is shown in figure
\ref{fig:5}, left panel, for the same simulations of figure
\ref{fig:4}. One sees that $\eps^2/\sigma^2(t)$ increases  monotonically
towards an asymptotic value,  again with a decreasing rate as $\eps$
is decreased, and eventually oscillates about it. 
The values of the equipartition times $\tau_{eq}(\eps)$ are reported in
logarithmic scale in figure \ref{fig:5}, right panel. Again, a
straight line with a slope not far from $-1.5$ is found to give a good
interpolation.

So the equipartition times $\tau_{eq}$ seem to follow a law of the form
\begin{equation}\label{eq:3}
  \tau_{eq}(\eps)= \frac {C'} {\eps^{3/2}} \  ,
\end{equation}
with a suitable $C'$.
Such a law seems to be in rather good qualitative agreement with the law
(\ref{eq:2}) for the relaxation time of the correlation of
polarization, but the
quantitative agreement is not that good. For example, at $\eps=295$
K the relaxation time for polarization was of only 0.5
picoseconds, while the relaxation time for equipartition  is  about 10 
picoseconds. 
 
In any case, the previous results seem to show  that the time needed for
time-averages to relax to their asymptotic values is larger than the
time needed for decorrelation to occur. 
 In the next subsection it will be seen that much larger  times are
 needed in order to  reproduce the spectra,  which means that the
attainment of equipartition doesn't flag the attainment of thermal 
equilibrium. This was already pointed out in the papers \cite{antonio}
(see Figure 6), \cite{danieli} (Figure 3) and  \cite{danieli2}, and
 discussed in \cite{moser}.
Eventually, this  fact will be strongly supported by the results
illustrated here in Section 4. 

\subsection{Relaxation times for the  spectra}
In the two  previous subsections we were  dealing  with indicators which 
 don't   correspond  to any real, physically observable, quantity.
As the two relaxation times there found 
 differ by at least one order of magnitude, it
is of interest to 
determine the relaxation time for a quantity that is physically
observable. So we deal here with reflectivity, the prototype of the measured
quantities in connection with spectra: we want to determine up to what
final time $t_{fin}$ should
an orbit be computed in order that  
reflectivity be determined in a reliable way. 
Let us recall that the reflectivity  $R(\omega)$ is 
connected  to the (complex) refractive index $n(\omega)$   
by the relation  (see Born's  handbook \cite{born})
$$
   R(\omega)= \left | \frac{1+n(\omega)}{1-n(\omega)} \right | \ 
$$
In turn, the refractive index $n(\omega)$  is   the square root of the electric 
permittivity $\epsilon(\omega)$,  which eventually  is determined  theoretically
 (see the next  section) from the Fourier transform of the correlation
$\overline{\vett P (t) \cdot \dot {\vett P}(0)}$, which  involves the
ionic polarization, i.e., the positions of the ions.

\begin{figure}
 \begin{center}
    \includegraphics[width = 1.\textwidth]{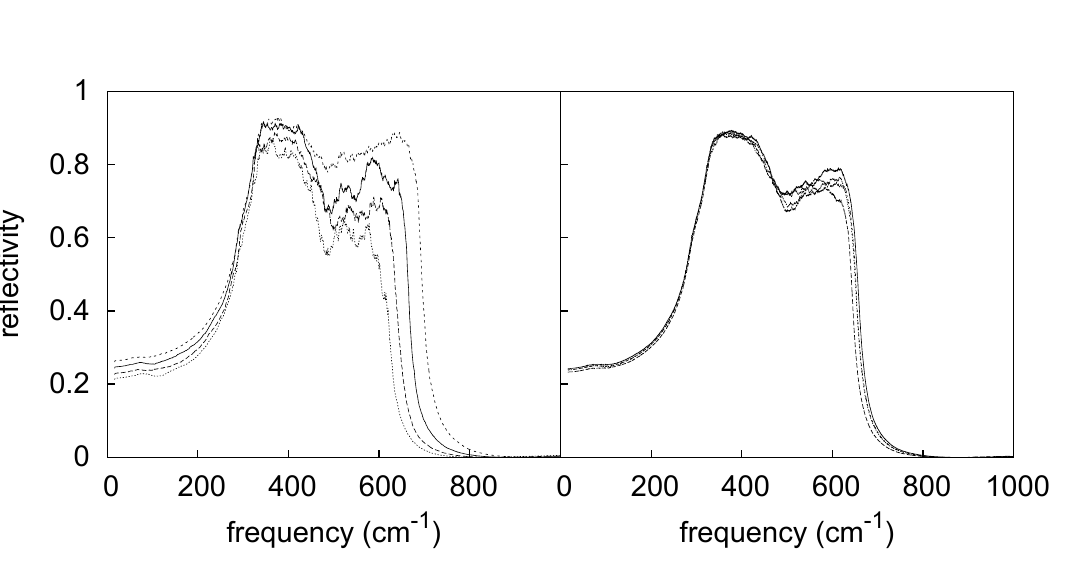}
  \end{center}\
  \caption{ Reflectivity curves computed for different stretches of a
    trajectory at $\eps=295$ K. Left panel: four curves corresponding to
    tracts of 40 picoseconds. Right panel: four curves corresponding
    to tracts of 400 picoseconds. Compare with Figure~\ref{fig:7},
    which involves an average over ten trajectories of 200
    picoseconds. $N=4096$.}
 \label{fig:6}
\end{figure}

So we would like   to know the time needed  for the time--average 
$\overline{\vett P (t) \cdot \dot {\vett P}(0)}$, and thus
reflectivity, to settle down to its asymptotic value.
One might  conjecture that this time coincides with the time required for
the variance $\sigma^2(t)$ of the 1 mode energies to attain
  its asymptotic value, i.e., its   Maxwell--Boltzmann value $\eps^2$.
For example, from figure~\ref{fig:5} (left panel) one sees that, 
for $\eps=295$ K, this time  is less than 40 picoseconds (perhaps being of the 
order of 10 picoseconds). So we take a
trajectory computed
for a sufficiently long time   $t_{max}$ (2
nanoseconds), and  subdivide it into stretches of length 
40 picoseconds. Then we compute the correlations
$\overline{\vett P (t) \cdot \dot {\vett P}(0)}$ as time--averages 
over any single stretch, thus  obtaining   distinct reflectivity
curves (i.e., values $R^i(\omega)$ for  for each $\omega$ in the domain
of interest). 

From figure~\ref{fig:6} one sees that such curves  differ
 greatly from each other for tracts of 40 picoseonds (left panel), 
and much less for tracts 
of 400 picoseconds (right panel), although a satisfactory stabilization 
was not yet attained. This 
means that such stretches  of the trajectory stick in different regions of
phase space, which actually give  rise to 
qualitatively different spectra. By trial and error we found that, at
$\eps= 295$ K, times of the order of 2 nanoseconds
are needed for the computed spectra to stabilize.
As the latter  time is larger than that needed
to attain equipartition, and much larger than the decay--time of the
correlation,  this result shows that the attainment
of equipartition doesn't correspond to the attainment of thermal
equilibrium. It appears that, notwithstanding the occurrence of
decorrelation and of equipartition,  partial times of 400 picoseconds
 are not sufficient for the partial
orbits  to explore the whole region actually   explored in 2 nanoseconds, which is
sufficient for the orbit to produce a spectrum agreeing  with the experimental one.

Now, by analogy with  the two relaxation times 
  previously investigated,  one is
naturally led to try to  estimate how the  stabilization time for reflectivity
too  depends on
specific energy $\eps$. However, unexpectedly, we met with 
serious difficulties in
implementing a  method providing  a reliable  quantitative
answer to this question. At first sight it seems that the latter time 
doesn't depend sensibly on specific energy $\eps$, but at the moment
we content ourselves with
the mentioned result, which refers to $\eps =295$ K.

In any case one is quite naturally  led to  conjecture that, in
analogy with what was found here in the two cases 
previously discussed, and was always found in the FPU
literature, 
also for the spectra the relaxation
 times should diverge as $\eps \to 0$. 
This, however, would make   a theoretical computation
 of the spectra  at low
temperatures impossible.
 In the next section we will show that, even if the relaxation times 
may diverge as $\eps \to 0$, a computation of the spectra is nevertheless
at hand, because the relaxation times
are actually bounded, keeping microscopic values. 
This is due to an unexpected fact,  
discovered in  paper \cite{physica}, which  concerns the identification 
of temperature in terms of specific energy.

\section{Behavior of the relaxation times at low temperatures}

So we illustrate how the problem of the divergence of the relaxation times
as $\eps\to 0$ is overcome, by making reference to the computation of a physical
quantity actually susceptible of comparison with experimental
data. i.e, the infrared spectrum, in its dependence on temperature $T$.
We summarize here results of our previous paper \cite{physica}.

The  theoretical quantity relevant  for the
spectra is the electric susceptibility tensor 
 $\chi_{ij}$ as a function of frequency  $\omega$ at a given 0
$T$. In the case of LiF, due to isotropy, one finds that it is sufficient 
 to deal with
its trace  $\chi=(1/3) \sum_i \chi_{ii}$.
The
contribution of the electrons to the permittivity
$\epsilon(\omega)$ in the infrared  can be
 taken into account through a constant $\chi^\infty$, so that  one finally has
\begin{equation*}
\epsilon(\omega)=1+4\pi \big[\chi^{ion}(\omega) + \chi^\infty\big] \ .
\end{equation*}
where $\chi^{ion}$ is the contribution due to the ions.
Through permittivity one can compute all quantities of interest as the
refractive index, the absorption coefficient and the reflectivity (see
the Born's handbook \cite{born}). We
0 on reflectivity, which is
 the quantity for which the largest set
of actually measured data is available.

In  the Green--Kubo approach 
the macroscopic response $\chi^{ion}(\omega)$ of the ions is
expressed by a formula which, in the semiclassical approximation,
at least formally reduces to the Fourier transform of a time--correlation 
involving  polarization, i.e., to the  formula 
\begin{equation}\label{kubo}
  \chi^{ion}(\omega, T) = \frac V{k_BT}\,  \,\int_0^{+\infty}
  e^{-i\omega t}\  \frac  {\langle
  \vett P (t) \cdot \dot {\vett P}(0)\, \rangle} 3\  \dif t \ ,
\end{equation}
in which the average  $ \langle \cdots  \rangle$ is   computed in principle
 according the classical Gibbs ensemble at temperature $T$. 

Now, in paper \cite{physica} we were unable to implement formula
(\ref{kubo}). The reason is that  the Gibbs measure is ill defined for
long--range potentials: one should  for example  impose a cutoff, and
then take the limit  in which the  cutoff tends to infinity. We chose
instead to perform the Ewald summation in which, by the way, energy is
somehow renormalized. The relations between such two methods is not
studied in the literature.

However, a  formula for the ionic susceptibility
  analogous  to (\ref{kubo}) can also be  deduced in a completely
  classical frame, 
as was actually done in 
 \cite{epj}. The only relevant  difference in such a  classical approach
is that any 
reference to ensembles is  avoided, as only time--averages are introduced. 
Indeed, the analogue of formula (\ref{kubo}) reads
\begin{equation}\label{kuboclassica}
  \chi^{ion}(\omega) = \frac V {\sigma^2_p}\ \int_0^{+\infty} e^{-i\omega t}
 \  \frac {\overline {\vett P(t) \cdot \dot{\vett  P}(0)}}3 \ \dif t \ ,
\end{equation}
where $\sigma^2_p$ is the variance of the momenta of the ions, and thus
 involves  their kinetic energy $K$, through  $\sigma^2_p=2K/3N$. Obviously,
 if the time--average coincides with the Gibbs average, one has    
$\sigma^2_p=k_BT$, and  formula (\ref{kuboclassica}) reduces exactly to  
(\ref{kubo}).
\begin{figure}
 \begin{center}
   \includegraphics[width= 1.\textwidth]{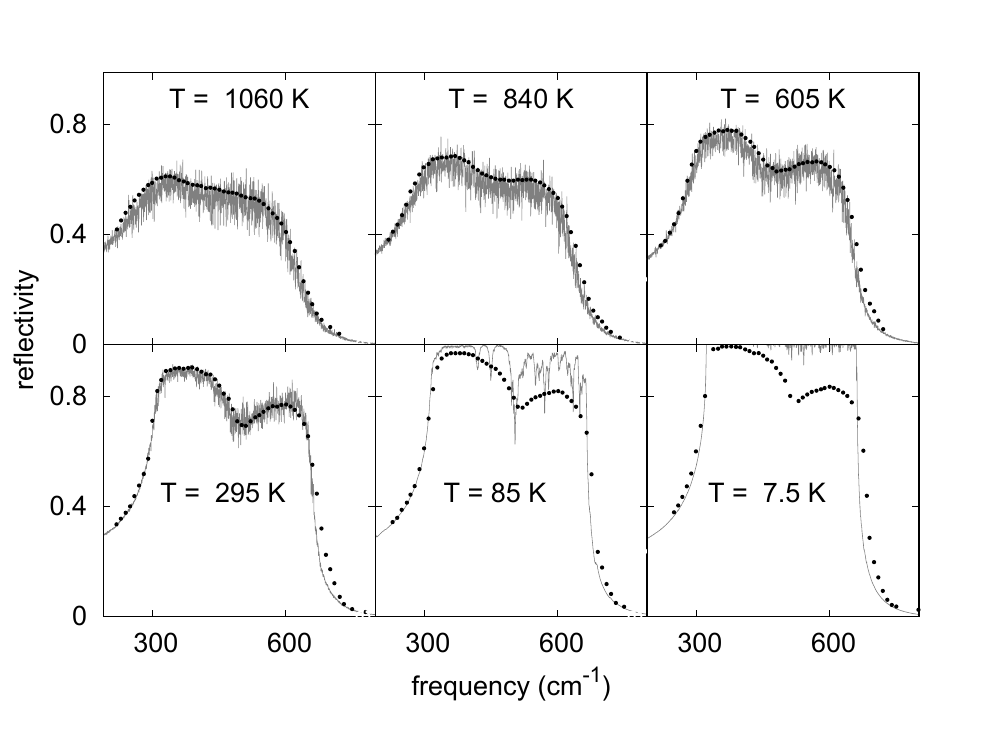}
 \end{center}
 \caption{ \label{fig:7} Reflectivity curves  of LiF  as a function 
of frequency at six
 temperatures. Results from calculations (solid line) are 
compared  with experimental data (points). Figure taken from
\cite{physica}. For any  temperature
$T$ the simulations were performed at a 0 energy 
 $\eps=k_BT$, and this leads to a disagreement at the two lower  
temperatures.  $N=4096$.}
\end{figure}

So, quite naturally we started   our computations by 
identifying temperature $T$ with kinetic energy
through the Gibbs prescription (\ref{eq:temperatura}),  i.e.,  
essentially through  $\eps= k_BT$.
The results of the numerical simulations are shown in figure \ref {fig:7},
where  the computed  spectra are reported together with the
experimental  data, for  six different  values of $T$. 

%

\begin{figure}
 \begin{center}
   \includegraphics[width= 1.\textwidth]{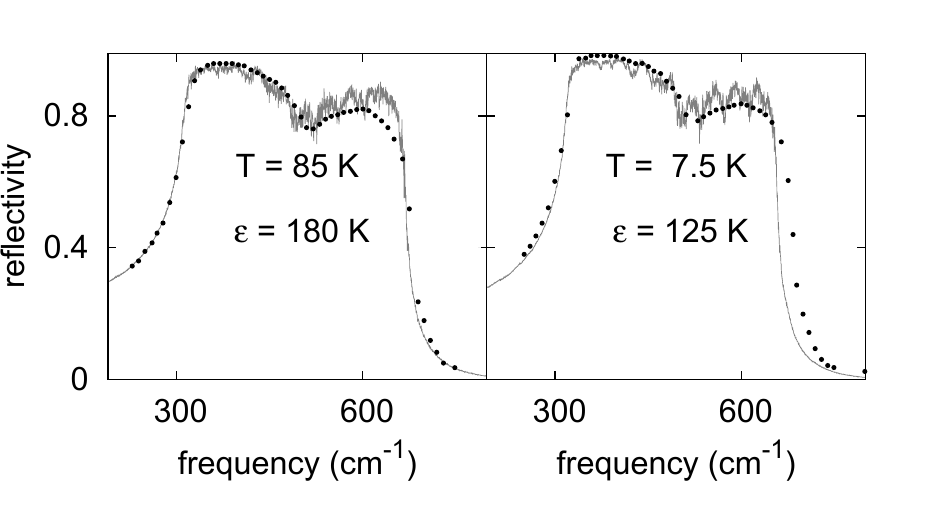}
 \end{center}
 \caption{ \label{fig:8} Same as Fig.~\ref{fig:7}, for the two
   temperatures 85~K and 7.5~K, with calculations performed 
   at specific energies  $\eps=180$~K and
   $\eps=125$~K, rather than 85 and 7.5 respectively.  $N=4096$.
}
\end{figure}

One sees that  
a good agreement is found at  room temperature and at
the higher ones,   actually
605~K, 840~K and 1060~K (just below melting). Instead, 
the agreement is seen to be  partially lost at 85~K, and even more so
at 7.5~K, where the secondary peak appears to be completely reabsorbed
in the main peak, the computed reflectivity sticking to the 
value 1.\footnote{This fact is quite peculiar. Indeed, at variance with the 
main peak, the frequency of the secondary one does not show up  among the 
normal--mode frequencies. So the secondary peak seems to be due to the 
nonlinearity, and its visibility should decrease with 0 energy,
whereas the contrary is seen to occur. }
However, it occurred to us to   find
that a pretty good agreement between experimental and computed spectra
 is recovered at the two low temperatures
of 85 K and  7.5 K too, if  one chooses  for $\varepsilon$ the values
180 K   and 125 K respectively.  This is shown in Fig.~\ref{fig:8}.

So, it is a fact that agreement is recovered at low temperatures
 if one renounces to stick at the Gibbs prescription.
This in particular implies that the specific energy $\varepsilon$ keeps  
a nonvanishing value even  at zero temperature, in qualitative
agreement with the experimental fact that a nonvanishing zero--point
energy exists. 
In turn, the nonvanishing of $\eps$ at vanishing temperature implies
that the relaxation times, 
which would diverge as $\varepsilon \to 0$, remain instead
 bounded in the  zero--temperature limit.
In other words, at low temperatures the classical theoretical formulas 
implemented through ionic  orbits obeying Newton's equations
 still apply,  reproducing pretty well the
experimental spectral curves, if one admits that Gibbs' statistical 
mechanics doesn't.

\section{Ergodicity properties}

Naturally,  it is universally accepted that classical Gibbs statistics
should be abandoned  at low temperatures, due to the increasing
relevance of purely
quantum phenomena, not describable in a classical frame. In such
a perspective the fact  that, at a  temperature as low  as 
7 K, in order to reproduce the spectra one has to work  at a value of  
specific energy $\varepsilon$  large  as 
about 125  K,  would be understood as corresponding to the existence of
a nonvanishing   zero--point energy. Nevertheless it seems that, apart
from such a correction,  classical dynamics is adequate to computing
the relevant correlations. One might perhaps 
describe such a situation through the slogan: ``At low temperatures
the dynamics is classical, while the statistics is a quantum
one''. The reason for such a behaviour remains however obscure.
 
Thus we preferred to follow, so to say, an ``internal'' approach, trying
to understand whether there exist reasons within classical physics
itself which allow one to ascertain whether Gibbs statistics is
justified  for our model or not. This clearly is a problem concerning
the ergodicity properties of our system, and so we presently start a
discussion of this point. See also the mathematical contributions
given by the Kozlov school \cite{kozlibro,koz1,koz2,koz3}. 

 First of all it is immediate to conclude
that,  by the fact itself of being  in a crystalline phase, 
the system is not ergodic.  This should be  obvious, as will be
recalled in  a moment, but in any case we exhibit
it here through an elementary counterexample. Namely,
using the Gibbs ensemble one finds that the phase--averaged particle
positions all have the same value, i.e., one has $<\vett x_j> = <\vett x_l>$ 
for all $j,l$. Indeed, as  the potential $\sum_{i,k}V(\vett x_k-\vett
x_i)$ is invariant under  the exchange $\vett x_j \leftrightarrow 
\vett x_l$, one has 
$$
\int_V \vett x_j \exp \big[-\beta \sum_{i,k}V(\vett
x_k-\vett x_i) \big] \dif \zeta =
\int_V \vett x_l \exp \big[-\beta \sum_{i,k}V(\vett
x_i-\vett x_k) \big] \dif \zeta \ ,
$$
However,  in the crystalline phase, for $j\neq l$ one has, for averages
computed over not too long times,
$\overline {\vett x}_j=\vett x^0_j\neq \vett {x}^0_l=\overline {\vett x}_l$,
where $\vett {x}^0_j$ and $\vett {x}^0_l$ are the lattice sites. So, 
there exist dynamical variables
with time averages differing from the phase averages, and  
the system is not ergodic.
\begin{figure}
 \begin{center}
   \includegraphics[width= 1.\textwidth]{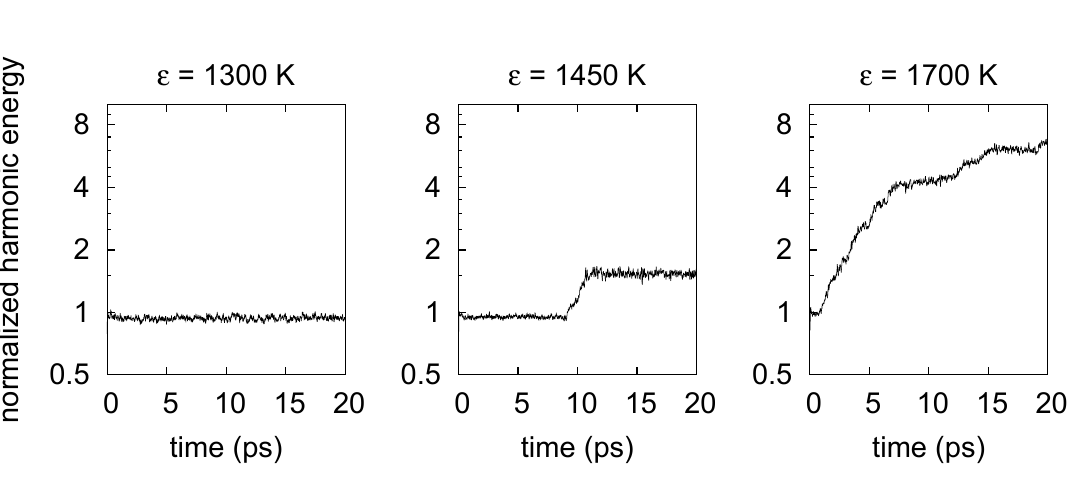}
 \end{center}
 \caption{ \label{fig:9} Normalized harmonic energy $H_{harm}/(H-E_{min})$ 
   versus  time for a time interval of 20 picoseconds, at
   $\eps=1300$ K, $\eps=1450$ K and $\eps=1700$  K, from left to
   right. Semilogarithmic scale. One passes from a fluctuation about 0.95, to a jump from 1
 to 1.6, and finally to a continuous increase from 1 to 7.  $N=512$
}
\end{figure}

A more physical reason for the system not being ergodic is that at low 
temperatures the
system's representative point in phase space remains confined in a
small region about the initial equilibrium point. This can be seen by
checking the value of $H_{harm}$,  
the second order truncation of the
expansion of $H$ about the equilibrium point. The function $H_{harm}$
 provides a
distance in phase space, since its level surfaces are ellipsoids,
centered about the equilibrium point. In fact, $H_{harm}$ is a positive
definite quadratic form in phase space, so that it can be put into
diagonal form by a familiar change of variables,   the new 
ones being exactly
the normal modes coordinates discussed in Section~2.2.

Figure~\ref{fig:9} (left) shows that, even at so large a specific
energy as $\eps =1300$ K the value of $H_{harm}$ remains bounded,
presenting very limited fluctuations, thus indicating that the point
is visiting a small region about the equilibrium point.  This fact is
immediately clear if one recalls that our model, as any
model of a crystal, actually presents $N!$ minima, which are obtained
from the particular one we considered up to now, by permuting in all
possible ways the particles. Moreover, such points are far
away from each other in phase--space, so that the corresponding values
of the function $H_{harm}$ (which refers to the expansion of the Hamiltonian $H$ about a
definite  particular minimum among the different $N!$ ones), takes
on very large values when evaluated about one of the remaining minima
as observed in an analogous case in \cite{amati}.  
On the other hand, for large enough $N$, at any
fixed $\eps$ no "kinematic" obstacle exists to the exchange of
adjacent particles, because the two particles
 just have to overcome a small finite
energy barrier. So, if the system were ergodic, for increasing time
one would observe a slow growth of the value of $H_{harm}$ until the value
$<H_{harm}>$ is attained, with a subsequent oscillation about it.

In fact the phenomenon of the ions diffusing along the lattice is
actually observed at sufficiently high values of $\eps$: for example
in the central panel of figure~\ref{fig:9}, which was obtained for
$\varepsilon=1450$ K, one can see a sudden jump of the value of $H_{harm}$,
which corresponds exactly, as could be visually exhibited,
 to an exchange of two adjacent particles of
the lattice.  However, as only two ions are involved, the phase point
did not yet wander along a large part of the  energy surface. Instead, in the
right panel of figure~\ref{fig:9}, which refers to $\varepsilon=1700$
K, one finds a continuous growth of $H_{harm}$: evidently the point is free
to wander in phase space, perhaps through the whole energy surface.
So, an ergodic behaviour may be conjectured to occur in the
latter case, but certainly not at lower energies: for
$\varepsilon<1300$ K one has $<H_{harm}> \neq {\overline H}_{harm}$, and the
system is not ergodic, \footnote{The exact calculation of $\langle
  H_{harm}\rangle$ is not feasible. However the property $\overline H_{harm}
  \neq \langle H_{harm}\rangle$ can be shown as follows: on the one hand
  one obviously has $\overline H_{harm} \simeq 3N \varepsilon$, on the other hand
  one has $\langle H_{harm}\rangle \simeq \frac 1{N!}\, \frac 1Z \sum_i
  \int_{U_i} H_{harm} \exp(-\beta H) d\zeta $ where $U_i$ is a suitable
  neighborhood of the $i$-th minimum, and the constant $Z$ the
  partition function.  Now, by symmetry the Gibbs weight is the same in all such
  neighborhoods, whereas $H_{harm}$ is much larger than
  $3N \varepsilon$ in all of them, apart from the initially chosen
  one.} at least up very long times. In a similar context it was shown
that the time required for the system to jump out of the  local
minimum
increases  exponentially as a function of $1/\eps$ (see \cite{ponno}, \cite{naumis}). 

The relevant point is that this happens notwithstanding the fact that 
equipartition of the normal--mode energies
had previously been attained.  Rather, it even turns out that the
occurring of equipartition is a signal that ergodicity does not hold.
Indeed, equipartition means $\overline {E_k}\simeq \eps$, whereas
ergodicity implies $\overline {E_k}\gg \eps$.
In conclusion: at low temperatures ergodicity does not hold up
to geological time scales, so that the Gibbs identification of
temperature in terms of kinetic energy per particle is not justified.

\section{Open problems}
We point out that, in fact,   the ergodic
problem is not really relevant for quantities of  a mechanical type, because
a comparison with experimental data involves time--averages rather than 
 phase--averages. This was amply illustrated in
the discussion following the Einstein contribution to the first Solvay
conference \cite {einstein}, and is  also currently done in  Molecular
Dynamics simulations.  The really hard problem is met when dealing
with quantities of thermodynamic type, namely,  temperature  and entropy,
because, for example, temperature is characterized as being an
integrating factor of the heat exchanged with another system, and not
as the average of some function in phase space. If expectations are
implemented as averages with respect to  the Gibbs measure, 
one  proves that the Lagrange
multiplier $\beta$ is the required integrating factor, which moreover
 coincides, in the classical case, with two
thirds the mean kinetic energy per particle.  If instead the
expectations are computed as time--averages, the problem of finding an
integrating factor is scarcely discussed in the literature, apart from
the classical works of Boltzmann and Clausius;
see however the recent paper \cite{carati}. So nobody
knows what the integrating factor is, nor whether it coincides with
the time--average of some definite dynamical variable.
\begin{figure}
 \begin{center}
   \includegraphics[width= 1.\textwidth]{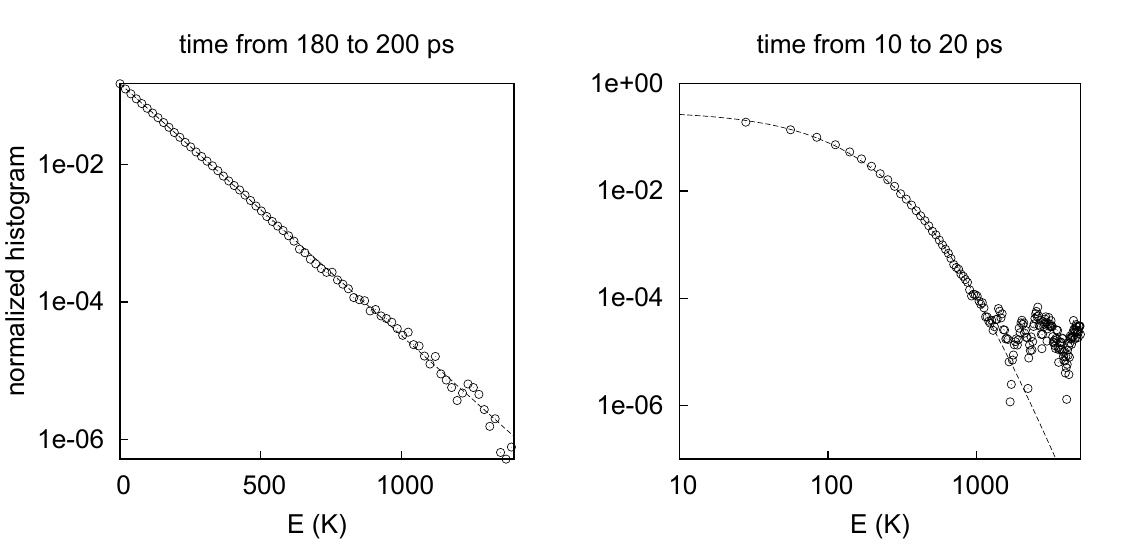}
 \end{center}
 \caption{ \label{fig:10} Maxwell--Boltzmann distribution and q--distribution. 
   Left panel. Histogram (in semilogarithmic scale) of the
   energies $E$ (circles) of the modes not initially excited, from time $t=180$
   (after their equipartition is attained) up to $t=200$ ps. Only the 15 modes
   of lowest frequency were initially excited, among the total  number 1536
   of modes.  Specific energy $\eps=120$ K, $N=512$.
   Solid line is the graph of the Maxwell--Boltzmann
   distribution function $C\exp(-E/\eps)$.
   Right panel. Same as left panel, in logarithmic scale, with data collected
   for time from 10 to 20 ps. Solid line is the graph of the Tsallis distribution
   function $C'(1+\beta (q-1)E)^{1/(1-q)}$ for $q=1.14$, $\beta^{-1}=67.8$ K.
   }
\end{figure}

It  seems that this problem could be solved if 
the expectations were implemented with respect to
 a suitable invariant phase--space measure (see for
example \cite{alberto}). So it  is of interest to understand
whether in our case it is possible  to find a measure which reproduces 
the actually computed time--averages. To this  end, following Kolmogorov,
 the first step  consists in determining the marginal distributions
and then checking their compatibility conditions. We just performed a 
preliminary work in this direction,  which consists in computing the
histograms of the mode energies, i.e. the marginals for a  single variable.
The results  are reported in figure~\ref{fig:10}, left panel,
in semilogarithmic scale, together with the fitting exponential curve 
(solid line), for  specific energy $\eps=120$ K and $N=512$.
The histogram was obtained as follows. One starts  from initial data of
FPU type, in the way previously illustrated (namely, with energy given  
only to the packet of  15   normal modes having the  lowest frequency),
and follows an orbit up to  180 picoseconds (a time 
 larger than the equipartition time: see figure~\ref{fig:5}, left panel). 
 Then one just considers the set of all normal modes that
 were not initially excited, and collects the values of the
normal mode energies for times from 180 up to 200 picoseconds.  One sees that the  
histogram agrees very well with the Maxwell-Boltzmann 
distribution $C\exp\, (-E/\eps)$.

We address now a further problem, that came to our attention after a
conversation  with   C. Tsallis. The problem is how are our researches
situated with respect to a thesis often discussed in recent works
(see for example 
\cite{tsallis1, tsallis2, tsallis3, rapisarda, ruffo, giansa}). Namely, 
the thesis that in systems with long--range 
interactions the  Maxwell--Boltzmann distribution should be attained 
only for very large  times, whereas  previously quasi stationary states would
be met, which present  ``nonclassical" distributions: moreover, the
crossover should take place at times which increase systematically as
the number $N$ of particles is increased. 
The problem is thus whether such a phenomenon is met 
in our model, or more in general in the whole domain of atomic
physics. Indeed, as particularly  emphasized by
Feynman,  the whole domain of atomic physics  is characterized by the 
presence of  long--range electromagnetic forces,
actually  in their retarded form,  
involving also the  radiation reaction force (see \cite{alessio} \cite{epj}).

  
This is indeed a very interesting question, that  we plan to address
more systematically in  future studies, and  here we limit ourselves 
to a preliminary result.  What we can presently say is that
the  above thesis may not be 
inconsistent with the phenomena occurring
in the realistic ionic crystal model. What we have  available at the moment
 is illustrated in figure 
 \ref{fig:10}, right panel, where we report    
an histogram obtained for the same run 
of the left panel, but with normal mode energies collected from 10 to
20 picoseconds.
Now, however, the scale is logarithmic    and the full curve 
is  that of the q--distribution $C\,[1-\beta (q-1)E]^{1/1-q}$ 
(with  $q= 1.14$ and 
$1/\beta=67.8$ K), which seems to fit pretty well the computed values.
This result shows  that, before the
 Maxwell--0 distribution is attained, a
q--distributions  is  exhibited  in our model too.
Whether a  persistence of the q--distribution for extremely long times  
 with increasing $N$ also occurs,  we cannot say at the moment, 
due to the difficulty of working with large numbers  of particles 
in our realistic three--dimensional model.

\section{Conclusions}
So we  have shown that the realistic ionic--crystal model we
investigated is not ergodic in the sense that, up to geological time
scales, the phase space trajectories explore only a very small part of
an energy surface. Thus, up to such times the use of the Gibbs measure
is not dynamically justified. Nevertheless the time--averages of physical
observables relax to asymptotic values within microscopic times, of
the order of nanoseconds. In particular this allows one to compute
theoretical infrared spectra, which moreover reproduce the
experimental spectra, even at extremely low temperatures.

This fact seems at first sight  to be in contrast with the 
picture one gets from the
numerical experiments performed on one--dimensional FPU models, in
which the relaxation times diverge as specific energy tends to
zero. Such a contradiction is overcome through the empirical
realization that temperature should not be identified with kinetic
energy per particle, as would be required by  Gibbs
statistical mechanics, if the latter were dynamically justified.
Finally, a  preliminary indication was given  that  
``nonclassical'' q-distributions may be significant for the realistic 
ionic--crystal  model too, and presumably for the whole domain  
of 0 physics.

How to formulate a consistent statistical mechanics, with the
corresponding suitable identification of temperature, in such
nonergodicity conditions is an open problem that seems to constitute
the modern form of the original FPU problem.


\section*{Acknowledgements}
Funding: research carried out with the support of resources of Big\&Open Data Innovation Laboratory (BODaI-Lab),
University of Brescia, granted by Fondazione Cariplo and Regione Lombardia.

\end{document}